\documentclass[journal]{IEEEtran}
\ifCLASSINFOpdf
\else
\fi
\usepackage[comma,numbers,square,sort&compress]{natbib}
\usepackage{algorithm} 
\usepackage{algorithmic} 
\usepackage{arydshln}
\usepackage{multirow} 
\usepackage{xcolor}
\usepackage{setspace}
\usepackage{amsmath}
\newtheorem{theorem}{\textbf{Theorem}}
\newtheorem{lemma}{\textbf{Lemma}}

\newtheorem{corollary}{\textbf{Corollary}}
\newtheorem{remark}{\textbf{Remark}}
\newtheorem{definition}{\textbf{Definition}}

\newtheorem{assumption}{\textbf{Assumption}}

\newenvironment{proof}{{\noindent{\bf \emph{Proof:}}}\quad}{\hfill $\square$\par}

\usepackage{multirow}
\usepackage{url}
\usepackage{subfigure}
\usepackage{fancyhdr}
\usepackage{amsmath}
\usepackage{multirow}
\usepackage{amssymb }
\usepackage{color}
\usepackage{graphics} 
\usepackage{graphicx}
\usepackage{geometry}

\renewcommand{\baselinestretch}{1.04} \normalsize

\begin{document}
	%
	%
	%
	%
	
	\title{\LARGE \bf
		Data-Driven Existence and Design of Target Output Controllers
	}
	
	%
	%
	
	\author{Yuan Zhang, Wenxuan Xu, Mohamed Darouach, Tyrone Fernando
		\thanks{Yuan Zhang and Wenxuan Xu are with School of Automation, Beijing Institute of Technology, Beijing, China. Mohamed Darouach is with the Centre de Recherche en Automatique de Nancy, Universit? de Lorraine, 54400 Cosnes et Romain, France. Tyrone Fernando is with the Department of Electrical, Electronic and
			Computer Engineering, University of Western Australia (UWA), Crawley, WA
			6009, Australia.  Email: zhangyuan14@bit.edu.cn, xuwenxuan1209@163.com, mohamed.darouach@univ-lorraine.fr, tyrone.fernando@uwa.edu.au. The work of the first two authors was supported by the
			National Natural Science Foundation of China under Grant 62373059. The work of the third author was supported by Gledden Senior Visiting
			Fellowship, Institute of Advanced Studies (IAS), UWA, Australia.}} 
	\maketitle

	\begin{abstract}Target output controllers aim at regulating a system's target outputs by placing poles of a suitable subsystem using partial state feedback, where full state controllability is not required. This paper establishes existence conditions for such controllers using input and partial state data, where the system dynamics are unknown. The approach bypasses traditional system identification steps and leverages the intrinsic structure of historical data to certify controller existence and synthesize a suitable feedback gain. Analytical characterizations are provided, ensuring that the resulting closed-loop system satisfies desired performance objectives such as pole placement or stabilization. Data-driven algorithms are then proposed to design target output controllers directly from data without identifying system parameters, where controllers with the order matching the number of target outputs and with minimum-order augmented target outputs are both addressed. Furthermore, a separation principle is revealed, decoupling the design of target output controllers from state observers. This enables the development of data-driven observer-based controllers that integrate estimation and control. Numerical examples validate the theoretical results and demonstrate the efficacy of the proposed approach.

	\end{abstract}
	

\section{Introduction}
For many engineered systems, such as autopilots in aerospace applications, full-state control is indispensable. By contrast, many biological, technological, and social systems exhibit such high dimensionality and complexity that controlling every state variable is neither feasible nor necessary \cite{gao2014target}. In these contexts, it suffices to regulate a linear functional of the state (or a subsystem) that is directly relevant to the system's transmission or task requirements; we refer to these quantities as target outputs \cite{van2017distance,czeizler2018structural}. A parallel scenario is estimating a linear functional of state rather than the whole state, and the so-called functional observers achieve this goal \cite{darouach2000existence,fernando2010functional}.

A dynamic system is said to be (target) output controllable if there exist external inputs capable of driving the specified target outputs from any initial to any desired final value within a finite time horizon \cite{chen1984linear}. Classical criteria for output controllability include rank conditions on the output controllability matrix, generalized Popov-Belevitch-Hautus (PBH) tests, and transfer-function-based conditions \cite{chen1984linear,fernando2025existence,montanari2024popov}. Graph-theoretic approaches have also elucidated how the interconnection structure influences output controllability \cite{van2017distance,li2020structural,zhang2025structural,zhang2023generic}. While these results guarantee the existence of point-to-point control laws for target outputs \cite{baggio2021data}, they do not systematically establish how a state-feedback controller can be constructed to achieve target output regulation regardless of initial or terminal conditions.

Recently, Fernando and Darouach \cite{fernando2025existence} addressed this gap by introducing the notion of a target output controller. Extending classical controllability concepts, they showed that by employing partial state feedback, one can assign the poles of a suitably defined subsystem to regulate the target outputs at an arbitrary convergence rate. Their work provided fundamental existence conditions and constructive design procedures, demonstrating that target output regulation remains achievable even in the absence of full-state controllability.

Despite these theoretical advances, the practical implementation of target output controllers is challenged by scenarios in which the underlying system dynamics are unknown. This is further complicated by the complexities involved in verifying the rank conditions necessary for the existence of such controllers.
Data-driven approaches have emerged as a promising alternative. Recent studies \cite{de2019formulas,coulson2019data,van2020data} have pioneered direct controller synthesis from experimental data, circumventing explicit model identification.
Notably, existing frameworks address linear quadratic regulation with full state data \cite{de2019formulas,van2020data}, predictive control via online optimization \cite{coulson2019data}, and point-to-point output regulation \cite{baggio2021data}.
However, these approaches either assume full-state accessibility and system controllability/stabilizability or do not address the synthesis of target output controllers.


Motivated by the potential of direct data-driven control paradigms, this paper presents a data-driven framework for the existence and design of target output controllers. Our key contributions are threefold. First, we establish necessary and sufficient conditions for the existence of target output controllers using only input and partial state data. This obviates the need for an explicit system model and yields a data-driven necessary condition for target output controllability, which is also sufficient in the class of diagonalizable systems. Second, we present methods for designing target output controllers directly from data, including the design of standard and augmented target output controllers. The latter introduces the flexibility of target augmentation to relax rank conditions, in which both model-based and data-driven characterizations for the minimum-order augmented target output controllers are provided.   Additionally, we explore a separation principle between the design of target output controllers and state observers, leading to the development of observer-based target output controllers. Several numerical examples demonstrate the effectiveness of the proposed approach.

The remainder of this paper is organized as follows. Section II reviews the necessary preliminaries and mathematical formulations. Section III presents the data-driven existence and design for target output controllers. Section IV explores the design of augmented target output controllers. Section V develops observer-based designs. Section VI illustrates the proposed methods with numerical simulations, and Section VII concludes the paper.

{\bf Notations.} We denote a row vector with entries $a_1,\dots,a_n$ by $(a_1,\cdots,a_n)$. For a matrix $M$, let $M^-$ be the generalized inverse of $M$ satisfying only the condition $MM^-M=M$. Note that if $M$ has full row rank, then $MM^-=I$, where $I$ denotes the identity matrix with a compatible dimension. Let $M^\top$  denote the transpose of matrix $M$, $\mathrm{eig}(M)$ the eigenvalues, and $\mathrm{rk}(M)$ the rank. We write $\mathrm{row}(M)$ for the number of rows of $M$. Finally, $\mathbb N$ denotes the set of nonnegative integers.

\section{Preliminaries} \label{section 2}


Consider a linear time-invariant system:
\begin{equation}\label{sys1}
\begin{gathered}
x(t+1)=A x(t)+B u(t), \\
y(t)=C x(t), \\
z(t)=F x(t),
\end{gathered}
\end{equation}
where $x(t) \in \mathbb{R}^{n}$ is the state variable, $y(t)\in \mathbb{R}^{p}$ is the measured output, $u(t)\in {\mathbb R}^{m}$ is the input, and $z(t)\in \mathbb{R}^{r}$ is a linear function of states.  $z(t)$ is called the target output, referring to linear functionals of states that are of interest.
{System \eqref{sys1} is said to be target output controllable if, for any initial target output $z_0\in {\mathbb R}^r$ and any final one $z_f\in {\mathbb R}^r$, there exists a $T>0$ and an input $u(t)$ defined on $\{0,1,\cdots,T-1\}$ that can drive $z(t)$ from $z(0)=z_0$ to $z(T)=z_f$.}  Without losing any generality, assume that $C$ and $F$ have full row rank, i.e., ${\rm rk}(F)=r$ and ${\rm rk}(C)=p$.

\begin{definition}[Target Output Controller, \cite{fernando2025existence}] A feedback controller $u(t)=KFx(t)$, $K\in {\mathbb R}^{m\times r}$, is called a target output controller of order $r$ if $u(t)$ can drive the target output $Fx(t)\rightarrow 0$ as $t\rightarrow\infty$ at an arbitrary rate of convergence from any initial condition $Fx(0)$ by placing $r$ poles of a subsystem of order $r$, that is, there exists a matrix $N\in {\mathbb R}^{r\times r}$ whose eigenvalues can be arbitrarily assigned such that $Fx(t)$ of system \eqref{sys1} follows the following dynamics
	\begin{equation}
	Fx(t+1)=F(Ax(t)+BKFx(t))=NFx(t).
	\end{equation}	
\end{definition}

Roughly speaking, a target output controller concerns regulating the target output $Fx(t)$ rather than the full state set $x(t)$, using local feedback information of the target output itself. This is appealing when controlling the full state is impossible or expensive, especially in large-scale network systems.

\begin{theorem}
	\citep[Theo 3]{fernando2025existence} \label{th1} The control law $u(t)=K F x(t)$, $K\in {\mathbb R}^{m\times r}$, is a $r$-order target output controller of system \eqref{sys1} if and only if the following two conditions hold
	\begin{equation}\label{modetest1}
	\begin{aligned}
	\mathrm{r k}\binom{FA}{F}=\mathrm{r k}(F) \\
	\end{aligned}
	\end{equation}
	\begin{equation}\label{modetest2}
	\begin{aligned}
	\mathrm{r k}(s F-F A \quad F B)=\mathrm{r k}(F), \forall s \in {\mathbb C} .
	\end{aligned}
	\end{equation}
	Moreover, when the above conditions hold, the poles of the target output controllers' are
	\begin{equation}\label{poles}
	{\text {eig}}\left(F A F^-+F B K\right).
	\end{equation}
\end{theorem}

\begin{lemma} \cite{fernando2025existence,montanari2024popov,zhang2023generic} \label{necessary}
	For system \eqref{sys1}, the following statements hold:
	\begin{itemize}
		\item[(1)] A necessary condition for the existence of a target output controller of order $r$ is that system \eqref{sys1} is target output controllable, which requires ${\rm rk}(F(B,AB,...,A^{n-1}B))=r$.
		\item[(2)] If system \eqref{sys1} is target output controllable, then condition \eqref{modetest2} holds.  
		\item[(3)] If $A$ is diagonalizable (in this case, we call system \eqref{sys1} diagonalizable), then system \eqref{sys1} is target output controllable if and only if condition \eqref{modetest2} holds.
	\end{itemize}
\end{lemma}

In Theorem \ref{th1}, if $F=I_n$, then conditions \eqref{modetest1} and \eqref{modetest2} reduce to exactly the controllability condition. On the other hand, even when system \eqref{sys1} is not controllable,
a target output controller can exist. 	The primary goal of this paper is to verify the existence conditions and design target output controllers directly from data.

\section{Data-driven existence and design of $r$-order target output controllers} \label{sec-3}
In this section, we provide data-driven existence conditions and design methods for $r$-order target output controllers. 
\subsection{Data Matrices}
Given a signal $w: {\mathbb N}\to {\mathbb R}^q$, denote $\{w(t)\}_{t=k}^{k+T}$ as the set
$\{w(k),w(k+1),...,w(k+T)\}$, and let
$W_{k:k+T}=\left(w^{\intercal}(k),...,w^{\intercal}(k+T)\right)^{\intercal}$, i.e., $W_{k:k+T}$ is the vectorized form of $w$ restricted to the interval $\{k,k+1,...,k+T\}$. Denote the Hankel matrix associated with $w$ as
{\small	\begin{equation*}
	W_{i, t, N}\!=\!\left(\begin{array}{cccc}
	w(i) & w(i+1) & \cdots &w(i+N-1)\\
	w(i+1) & w(i+2) & \cdots& w(i+N)\\
	\vdots &\vdots  & \vdots &\vdots  \\
	w(i+t-1) & w(i+t) & \cdots &w(i+t+N-2)
	\end{array}\right)
	\end{equation*}}where the first subscript $i$ is the time of the first sample of the signal in this Hankel matrix, the second one $t$ is the number of row blocks, and the third one $N$ is the number of columns. Specially, if $t=1$, we simply denote $W_{i,1,N}$ by $W_{i,N}$, i.e., $W_{i,N}=\left(w(i),...,w(i+N-1)\right)$.

\begin{definition}
	The sequence $W_{k:k+T}$ is said to be persistently exciting of order $t$ if $W_{k,t,T+2-t}$ has full row rank.
\end{definition}

Let $U=\left\{u_d(t)\right\}_{t=0}^{T}$ and $Y=\left\{y_d(t)\right\}_{t=0}^{T}$ be the input-output data of system \eqref{sys1} collected during an experiment (called historical data). Let $X=\left\{x_d(t)\right\}_{t=0}^{T}$ and $Z=\left\{z_d(t)\right\}_{t=0}^{T}$ be the state data and target output associated with $U$ and $Y$. That is, each tuple $(x_d(t+1),x_d(t),u_d(t),y_d(t),z_d(t))$ satisfies \eqref{sys1} with $t=0,...,T$. Moreover, we divide each signal into the past part and the future part as follows
$$
\begin{aligned}
& U_p=\left(u_d(0), \cdots, u_d(T-1)\right),Y_p=\left(y_d(0), \cdots, y_d(T-1)\right),\\
& X_p=\left(x_d(0), \cdots, x_d(T-1)\right), Z_p=\left(z_d(0), \cdots, z_d(T-1)\right),\\
& U_f=\left(u_d(1), \cdots, u_d(T)\right), Y_f=\left(y_d(1), \cdots, y_d(T)\right), \\
& X_f=\left(x_d(1), \cdots, x_d(T)\right), Z_f=\left(z_d(1), \cdots, z_d(T)\right).
\end{aligned}
$$Here, the subscript $p$ denotes the past sequence, and $f$ the future sequence. 
Namely, $W_p=W_{0,T}$, $W_f=W_{1,T}$, with $W=U,Y,X,Z$. It is easy to verify that
\begin{equation} \label{linear-relation}
\begin{aligned}
X_f&=AX_p+BU_p,\\
Y_p&=CX_p,\\
Z_p&=FX_p.
\end{aligned}
\end{equation}
We make the following assumption on the historical data.
\begin{assumption}[Persistent Excitation of Data]\label{as1}
	The historical data satisfies that $\binom{U_p}{X_p}$ has full row rank.
\end{assumption}

\begin{remark}
	The above assumption on the persistent excitation (PE) of data is common in the literature on direct data-driven control \cite{de2019formulas} and system identification \cite{verhaegen2007filtering}. The full row rank of $\binom{U_p}{X_p}$ implies that the state vector is sufficiently excited and there is no linear feedback from the states to the inputs.
	To ensure this condition, the data length $T\ge n+m$.
	Even when the full state data $X$ is not available, this assumption can be satisfied by choosing sufficiently random input and sufficiently random initial state during an experiment \cite{verhaegen2007filtering}. In particular, if system \eqref{sys1} is controllable and the input $U_{0:T}$ is persistently exciting of order $n+1$, then Assumption \ref{as1} is
	guaranteed~\cite{van2020willems}.
\end{remark}

The following lemma on the rank of the product of two matrices will be frequently used in our derivations.
\begin{lemma}\citep[Lem 2.1]{verhaegen2007filtering}\label{lem1}
	Given matrices $M$ and $N$ with compatible dimensions, if $N$ has {\emph{full row rank}}, then ${\rm rk}(MN)={\rm rk}(M)$.
\end{lemma}
\subsection{Existence and Design of $r$-Order Target Output Controllers}	
{
	\begin{theorem}[Data-Driven Existence Condition]\label{th2}
		Under Assumption~\ref{as1}, a target output controller of order $r$ exists if and only if
		\begin{equation}\label{modetest4}
		\begin{array}{ll}
		\mathrm{r k}\left(\begin{array}{c}
		U_p \\
		Z_p \\
		Z_f
		\end{array}\right)=\mathrm{r k}\left(\begin{array}{c}
		U_p \\
		Z_p
		\end{array}\right), \\
		\end{array}
		\end{equation}
		\begin{equation}\label{modetest5}
		\begin{array}{ll}
		\mathrm{r k}\left(\lambda Z_p-Z_f\right)=\mathrm{r k}\left(Z_p\right), & \forall \lambda \in {\mathbb C}.
		\end{array}
		\end{equation}
	\end{theorem}
	
	\begin{proof}
		Observe that
		$$
		\left(\begin{array}{c}
		U_p \\
		Z_p
		\end{array}\right)=\left(\begin{array}{c}
		U_p \\
		F X_p
		\end{array}\right)=\left(\begin{array}{ll}
		0 & I_m \\
		F & 0
		\end{array}\right)\left(\begin{array}{c}
		X_p \\
		U_p
		\end{array}\right)
		$$
		Under Assumption~\ref{as1}, Lemma \ref{lem1} yields  $$\quad \mathrm{{rk}}\left(\begin{array}{c}U_p \\ Z_p\end{array}\right)={\rm rk}\left(\begin{array}{ll}
		0 & I_m \\
		F & 0
		\end{array}\right)=\mathrm{{rk}}(F)+m.$$
		Moreover, observe that
		{\small{	$$
				\left(\begin{array}{c}
				U_p \\
				Z_p \\
				Z_f
				\end{array}\right)=\left(\begin{array}{c}
				U_p \\
				F X_p \\
				F\left(A X_p+B U_p\right)
				\end{array}\right)=\left(\begin{array}{cc}
				0 & I \\
				F & 0 \\
				F A & F B
				\end{array}\right)\left(\begin{array}{c}
				X_p \\
				U_p
				\end{array}\right)
				$$}}
		Again, Lemma \ref{lem1} yields
		$$\mathrm{r k}\left(\begin{array}{c}
		U_p \\
		Z_p \\
		Z_f
		\end{array}\right)=\mathrm{r k}\left(\begin{array}{cc}
		0 & I \\
		F & 0 \\
		F A & F B
		\end{array}\right)=\mathrm{r k}\left(\begin{array}{c}
		F \\
		F A
		\end{array}\right)+m.
		$$
		Therefore, condition \eqref{modetest1} is equivalent to condition \eqref{modetest4}.
		Next, note that
		$$ 	\begin{aligned}
		\lambda Z_p-Z_f&=\lambda F X_p-F\left(A X_p+B U_p\right)\\
		&=\left(\begin{array}{cc}
		\lambda F-F A & -F B \\
		\end{array}\right)\left(\begin{array}{c}
		X_p \\
		U_p
		\end{array}\right), \forall \lambda\in {\mathbb C}.
		\end{aligned}
		$$
		From Lemma \ref{lem1}, we have
		$\mathrm{r k}\left(\lambda Z_p-Z_f\right)=\mathrm{r k}\left(\begin{array}{cc}
		\lambda F-F A & F B \\
		\end{array}\right)$.
		Since
		$
		\mathrm{r k}\left(Z_p\right)=\mathrm{r k}\left(F X_p\right)=\mathrm{r k}(F)
		$ with Assumption \ref{as1}, we obtain that
		condition \eqref{modetest2} is equivalent to condition~\eqref{modetest5}.
	\end{proof}
	
	Based on Lemma \ref{necessary}, we have the following data-based necessary condition for target output controllability.
	
	\begin{corollary}\label{corollary-2}  Under Assumption \ref{as1}, system \eqref{sys1} is target output controllable only if condition \eqref{modetest5} holds. Moreover, if system \eqref{sys1} is diagonalizable, this condition is also sufficient.
	\end{corollary}
	
	\begin{proof}
		Based on Lemma \ref{necessary}, the result follows from the proof of Theorem \ref{th2}.
	\end{proof}
	
	\begin{remark}
		It is yet unclear how to check the diagonalizability of $A$ directly from data without identifying it. On the other hand, it is indicated in \cite{zhang2023generic} that a randomly generated matrix (with random zero-nonzero patterns and random values for each nonzero entry) of dimension $n$ is
		diagonalizable almost surely as $n$ approaches infinity. This implies that condition \eqref{modetest5} may be a necessary and sufficient condition for output controllability in most practical cases.
	\end{remark}

	We say an asymptotic target output controller of order $r$ exists if there is a Schur stable $N\in {\mathbb R}^{r\times r}$ such that $Fx(t+1)=F(Ax(t)+BKFx(t))=NFx(t)$. This means $F x(t) \rightarrow 0$ as $t \rightarrow \infty$ by the control law $u(t)=KFx(t)$, $\forall x(0)\in {\mathbb R}^n$.  From the proof of Theorem \ref{th1} in \cite{fernando2025existence}, we have the following corollary.
	
	\begin{corollary}\label{co1}
		An asymptotic target output controller of order $r$ exists if and only if
		\begin{equation}\label{modetest6}
		\mathrm{r k}\binom{F A}{F}=\mathrm{r k}(F), \\
		\end{equation}
		\begin{equation}\label{modetest7}
		\mathrm{r k}(\lambda F-F A \quad F B)=\mathrm{r k}(F), \quad \forall \lambda \ {\rm with} \ |\lambda| \geqslant 1 .
		\end{equation}
	\end{corollary}
	
	\begin{corollary}
		\label{th3}
		Under Assumption~\ref{as1}, an asymptotic controller of order $r$ exists if and only if
		\begin{equation}\label{datatest8}
		\mathrm{r k}\left(\begin{array}{c}
		U_p \\
		Z_p \\
		Z_f
		\end{array}\right)=\mathrm{r k}\left(\begin{array}{c}
		U_p \\
		Z_p
		\end{array}\right),
		\end{equation}
		\begin{equation}\label{datatest9}
		\mathrm{r k}\left(\lambda Z_p-Z_f\right)=\mathrm{r k}\left(Z_p\right), \quad \forall \lambda \text { with }|\lambda| \geqslant 1 .
		\end{equation}
	\end{corollary}

	\begin{theorem}[Alternative Existence Conditions]\label{th4}
		Under Assumption \ref{as1}, the following three conditions are equivalent
		
		(1)
		\begin{equation}\label{modetest4a}
		\begin{array}{ll}
		\mathrm{r k}\left(\begin{array}{c}
		U_p \\
		Z_p \\
		Z_f
		\end{array}\right)=\mathrm{r k}\left(\begin{array}{c}
		U_p \\
		Z_p
		\end{array}\right), \\
		\end{array}
		\end{equation}
		\begin{equation}\label{modetest5a}
		\begin{array}{ll}
		\mathrm{r k}\left(\lambda Z_p-Z_f\right)=\mathrm{r k}\left(Z_p\right), & \forall \lambda \in {\mathbb C}.
		\end{array}
		\end{equation}
		
		(2)
		$Z_f$ can be written as
		{  \begin{equation} \label{T1T2}
			Z_f=\left(\begin{array}{ll}T_1 & T_2\end{array}\right)\left(\begin{array}{c}U_p \\ Z_p\end{array}\right),\end{equation}}where $T_2$ is square, such that $\left(T_2, T_1\right)$ is controllable.
		
		(3) Condition \eqref{modetest4a} holds and $\left(Z_fZ_p^-, Z_f(I-Z_p^-Z_p)\right)$ is controllable.
	\end{theorem}
	
	\begin{proof}
		$(2) \Rightarrow(1):$ From $Z_f=T_1 U_p+T_2 Z_p$, we have \eqref{modetest4a} holds. Moreover, for each $\lambda\in {\mathbb C}$,	
		$$
		\begin{aligned}
		& \mathrm{r k}\left(\lambda Z_p-Z_f\right) \\
		= & \mathrm{r k}\left(\lambda Z_p-T_1 U_p-T_2 Z_p\right) \\
		= & \mathrm{r k}\left(\begin{array}{ll}
		\lambda I-T_2 & T_1
		\end{array}\right)\left(\begin{array}{c}
		Z_p \\
		-U_p
		\end{array}\right) \\
		= & \mathrm{r k}\left(\begin{array}{ll}
		\lambda I-T_2 & T_1
		\end{array}\right),
		\end{aligned}
		$$where the last equality comes from Lemma \ref{lem1} and the fact that $\left(\begin{array}{c}
		Z_p \\
		U_p
		\end{array}\right)=\left(\begin{array}{cc}
		F & 0 \\
		0 & I
		\end{array}\right)\left(\begin{array}{c}
		X_p \\
		U_p
		\end{array}\right)$ has full row rank.
		As $\left(T_2, T_1\right)$ is controllable, ${\rm rk}\left(\lambda I-T_2 \quad T_1\right)=r$, implying that $\mathrm{r k}\left(\lambda Z_p-Z_f\right)=r$,
		$\forall \lambda \in {\mathbb C}$.
		
		$(1) \Rightarrow(2)$ : Condition \eqref{modetest4a} implies that there is a pair $\left(T_1, T_2\right)$ such that $Z_f=T_1 U_p+T_2 Z_p$.
		Since $\binom{Z_p}{U_p}$ has full row rank, condition \eqref{modetest5a} yields $$
		\begin{aligned}
		&\mathrm{ r k}\left(\lambda Z_p-Z_f\right)
		=  \mathrm{r k}\left(\lambda Z_p-T_1 U_p-T_2 Z_p\right) \\
		= & \mathrm{r k}\left(\begin{array}{ll}
		\lambda I-T_2 & T_1
		\end{array}\right)\left(\begin{array}{c}
		Z_p \\
		-U_p
		\end{array}\right)
		=  \mathrm{r k}\left(\begin{array}{ll}
		\lambda I-T_2 & T_1
		\end{array}\right) \\
		=& r, \quad \forall \lambda \in {\mathbb C},
		\end{aligned}
		$$
		which implies the controllability of $\left(T_2, T_1\right)$.
		
		(1) $\Leftrightarrow$ (3): It suffices to prove that condition (\ref{modetest5a}) is equivalent to the controllability of $\left(Z_fZ_p^-, Z_f(I-Z_p^-Z_p)\right)$. Since $(Z_p^-, I-Z_p^-Z_p)$ has full row rank, for any $\lambda \in {\mathbb C}$, we have
		$$\begin{aligned}
		{\rm rk}(\lambda Z_p-Z_f)&={\rm rk}\left((\lambda Z_p-Z_f)(Z_p^-, I-Z_p^-Z_p)\right)\\
		&={\rm rk}\left((\lambda I-Z_fZ_p^-,-Z_f(I-Z_p^-Z_p))\right),
		\end{aligned}$$where $Z_pZ_p^-=I_r$ has been used (noting that $Z_p$ has full row rank).
		The proof is then finished by the PBH test for controllability.
	\end{proof}
	
	{Note that $\left(\begin{array}{c}
		U_p \\
		Z_p
		\end{array}\right)$ has full row rank. From \eqref{T1T2}, it follows that $\left(T_1\ \ T_2\right)$, if exists (namely, condition \eqref{modetest4a} is satisfied), can be uniquely\footnote{From \eqref{T1T2}, we know that under Assumption \ref{as1}, $(T_1,T_2)$ is unique if exists. Moreover, $\left(\begin{array}{c}
			U_p \\
			Z_p
			\end{array}\right)\left(\begin{array}{c}
			U_p \\
			Z_p
			\end{array}\right)^-=I$ gives that $(T_1,T_2)$ can be expressed as $Z_f\left(\begin{array}{c}
			U_p \\
			Z_p
			\end{array}\right)^-$.} written as $\left(T_1\ \ T_2\right)=Z_f\left(\begin{array}{c}
		U_p \\
		Z_p
		\end{array}\right)^-$.} Compared to Theorem \ref{th2},  condition (2) in Theorem \ref{th4} is not only computationally more efficient, but also provides a practical procedure to design a target output controller with order $r$, as shown in the following theorem.
	\begin{theorem}[Data-Driven Target Output Controller Design]\label{th5}
		Suppose the conditions in Theorem~\ref{th2} hold under Assumption~\ref{as1}. Let $T_1,T_2$ be defined as in Theorem~\ref{th4} and  $K$ be such that $T_1 K+T_2$ has a desired set of poles $\Lambda$. Then, the feedback controller $u(t)=K F x(t)$ is a target output controller with the pole set $\Lambda$. The closed-loop dynamics of the $r$-order subsystem corresponding to the target output is
		$$
		z(t+1)=\left(T_1 K+T_2\right) z(t),
		$$
		where $z(t)=Fx(t)$.
	\end{theorem}
	
	\begin{proof}
		With the control law $u(t)=K F x(t)=K z(t)$, the closed-loop dynamics of system \eqref{sys1} is
		$$
		\begin{aligned}
		x(t+1) & =A x(t)+B K F x(t)
		=(A+B K F) x(t),
		\end{aligned}
		$$
		leading to $$F x(t+1)=F(A+B K F) x(t).$$
		Since $\left(\begin{array}{c}U_p \\ X_p\end{array}\right)$ has full row rank, let $G_K$ be such that
		$\left(\begin{array}{c}
		K F \\
		I_n
		\end{array}\right)=\left(\begin{array}{c}
		U_p \\
		X_p
		\end{array}\right) G_K$, yielding
		${\left(\begin{array}{c}
			K F \\
			F
			\end{array}\right)=\left(\begin{array}{c}
			U_p \\
			Z_p
			\end{array}\right) G_K}$.
		We then have $$\begin{aligned}F(A+B K F) x(t)&=F\left(\begin{array}{ll}
		B & A
		\end{array}\right)\left(\begin{array}{c}K F \\ I_n\end{array}\right) x(t)\\
		& =F\left(\begin{array}{ll}
		B & A
		\end{array}\right)\left(\begin{array}{c}
		U_p \\
		X_p
		\end{array}\right) G_K x(t) \\
		& =F X_f G_K x(t)
		\end{aligned}
		$$
		Since $F X_f=\left(\begin{array}{ll}
		T_1 & T_2
		\end{array}\right)\left(\begin{array}{c}
		U_p \\
		Z_p
		\end{array}\right),
		$ we have
		$$
		\begin{aligned}
		F X_f G_K x(t) & =\left(\begin{array}{ll}
		T_1 & T_2
		\end{array}\right)\left(\begin{array}{c}
		U_p \\
		Z_p
		\end{array}\right) G_K x(t) \\
		& =\left(\begin{array}{ll}
		T_1 & T_2
		\end{array}\right)\left(\begin{array}{c}
		K F \\
		F
		\end{array}\right) x(t) \\
		& =\left(\begin{array}{ll}
		T_1 & T_2
		\end{array}\right)\left(\begin{array}{c}
		K \\
		I_n
		\end{array}\right) F x(t) \\
		&=\left(T_1 K+T_2\right) z(t).
		\end{aligned}
		$$
		The above two equalities yield
		$$\quad z(t+1)=F(A+B K F) x(t)=\left(T_1 K+T_2\right) z(t).
		$$
		Since $(T_2,T_1)$ is controllable via Theorem \ref{th4}, $T_1K+T_2$ can have an arbitrary set of poles by choosing $K$.
	\end{proof}
	Theorem \ref{th5} reveals that the design of target output controllers can be reduced to a pole assignment problem, with all the matrices involved extracted purely from the data. Note that the existence conditions in Theorem \ref{th4} and the design approach in Theorem \ref{th5} require only the historical data $U$ and $Z$, which is generically not sufficient to identify the system parameters $(A,B,F)$ uniquely. This highlights the advantage of the proposed data-driven approach when the system is not identifiable from the available data.
	
	\section{Controller design by target augmentation}
	When conditions \eqref{modetest1}-\eqref{modetest2} are not satisfied, \cite{fernando2025existence} proposed that these conditions can be relaxed by introducing some additional target output $Rx(t)\in {\mathbb R}^{n_0-r}$ so that conditions \eqref{modetest1}-\eqref{modetest2} are satisfied with $F$ replaced by the new augmented full row rank target output matrix $\binom{F}{R}$, where $r<n_0\le n$. The reasoning of doing so is reflected by the following lemma.
	
	\begin{lemma} \cite{fernando2025existence} \label{modelaug}
		The control law $u(t)=K\binom{F}{R}x(t)$, where $R\in {\mathbb R}^{(n_0-r)\times n}$ and $K\in {\mathbb R}^{m\times n_0}$, can drive the functional $\binom{F}{R}x(t)\to 0$ as $t\to \infty$ at an arbitrary rate from any initial condition $\binom{F}{R}x(0)$ by placing $n_0$ poles of a subsystem of order $n_0$ if and only if
		conditions \eqref{modetest1}-\eqref{modetest2} are  satisfied with $F$ replaced by $\binom{F}{R}$.
	\end{lemma}
	
	If the conditions in Lemma \ref{modelaug} hold, we can find a target output controller of order $n_0$ that can arbitrarily regulate $\binom{F}{R}x(t)$ to zero. We call such a controller by the {\emph{augmented target output controller of order $n_0$}}.
	
	Based on Lemma \ref{modelaug} and Theorem \ref{th2}, it is readily seen that the data-based existence condition of an augmented target output controller can be stated as follows.
	
	\begin{lemma}
		Under Assumption \ref{as1}, an augmented target output controller of order $n_0$ exists if and only if there exists a matrix $R\in {\mathbb R}^{(n_0-r)\times n}$ such that
		\begin{equation}\label{datatest10aug}
		\mathrm{r k}\left(\begin{array}{c}
		U_p \\
		Z_p \\
		R X_p
		\end{array}\right)=\mathrm{r k}\left(\begin{array}{c}
		U_p \\
		Z_p \\
		R X_p \\
		Z_f \\
		R X_f
		\end{array}\right), \ {\rm and}
		\end{equation}
		\begin{equation}\label{datatest5aug}
		\begin{array}{ll}
		\mathrm{r k}\left(\lambda \binom{Z_p}{RX_p}-\binom{Z_f}{RX_f}\right)=n_0, & \forall \lambda \in {\mathbb C}.
		\end{array}
		\end{equation}
	\end{lemma}
	
	Note that condition \eqref{datatest5aug} implies condition \eqref{modetest5}. Therefore, condition \eqref{modetest5} is a necessary condition for the existence of an augmented target output controller of order $n_0$ for any $n_0\ge r$.
	
	Under Assumption \ref{as1}, a similar argument to Theorem \ref{th4} leads to that once condition \eqref{datatest10aug} holds, condition \eqref{datatest5aug}  can be checked by first finding matrices  $T_1,T_2$ such that $\binom{Z_f}{RX_f}=T_1U_p+T_2\binom{Z_p}{RX_p}$ and then checking the controllability of $(T_2,T_1)$. Note that controllability is a generic property in the sense that if $(T_2,T_1)$ is controllable, then $(T_2+\delta T_2, T_1+\delta T_1)$ is also controllable for almost all
	$\delta T_2 \in {\mathbb R}^{n_0\times n_0}$ and $\delta T_1 \in {\mathbb R}^{n_0\times m}$ \cite{Lin_1974}. It might be safe to consider condition \eqref{datatest5aug} as a secondary condition with respect to condition \eqref{datatest10aug}.
	
	From the proof of Theorem \ref{th2}, it is readily seen that under Assumption~\ref{as1}, condition \eqref{datatest10aug} holds if and only if condition \eqref{modetest1} holds with $F$ replaced by $\binom{F}{R}$ for some $R\in {\mathbb R}^{n_0-r}$, i.e.,
	\begin{equation}\label{rank-condition}
	\begin{aligned}{\rm rk}
	\left(\begin{array}{c}
	FA \\
	RA \\
	F\\
	R
	\end{array}\right)={\rm rk}\left(\begin{array}{c}
	F\\
	R
	\end{array}\right).
	\end{aligned}
	\end{equation}
	

	{In what follows, we first give a model-based analytical solution for the minimum $n_0$ satisfying condition \eqref{rank-condition}, which is inspired by \citep[Theo 6]{fernando2025existence}. We then provide a data-driven method to construct such a $R$ with the minimum number of rows satisfying \eqref{datatest10aug}. Let $\mathrm{O}_t(A,F)$ be the $t$ - step observability matrix of $(A,F)$, i.e.,  \begin{equation*}
		O_t(A,F)=\left(\begin{array}{c}
		F \\
		F A \\
		\vdots \\
		F A^{t-1}
		\end{array}\right).
		\end{equation*}
		Denote the observability matrix of $(A,F)$ by $O(A,F)$, i.e., $O(A,F)=O_{n}(A,F)$.
		
		\begin{theorem} \label{thlower}
			Let $R\in {\mathbb R}^{n_0-r}$ be a matrix such that \eqref{rank-condition} holds. Then, ${\rm row}(R)\ge d_{\min}\doteq {\rm rk}(O(A,F))-r$. Moreover, let $R$ consist of $d_{\min}$ rows of $O(A,F)$ such that ${\rm rk}\left(\begin{array}{c}F\\R \end{array}\right)={\rm rk}(O(A,F))$. Then, $R$ is a matrix with the minimum number of rows that satisfies \eqref{rank-condition}.
		\end{theorem}
		
		\begin{proof}
			Let $d={\rm rk}(O(A,F))$. If $R$ satisfies \eqref{rank-condition}, we have
			\begin{equation}\label{inequality} {\rm rk}(O(A,\binom{F}{R}))={\rm rk}
			\left(\begin{array}{c}
			FA \\
			RA \\
			F\\
			R
			\end{array}\right)={\rm rk}\left(\begin{array}{c}
			F\\
			R
			\end{array}\right).\end{equation}
			Since ${\rm rk}(O(A,\binom{F}{R}))\ge {\rm rk}(O(A,F))$, from \eqref{inequality} we have
			$${\rm row}(R)\ge {\rm rk}(O(A,\binom{F}{R}))-{\rm rk}(F)\ge d-r.$$Therefore, $d_{\min}$ is a lower bound of ${\rm row}(R)$.
			On the other hand, suppose that $R$ is constructed in the way stated in Theorem \ref{thlower}. Note that
			$$\begin{aligned}{\rm rk}
			\left(\begin{array}{c}
			FA \\
			O(A,F)A \\
			F\\
			O(A,F)
			\end{array}\right)={\rm rk}(O(A,F))={\rm rk}\left(\begin{array}{c}
			F\\
			O(A,F)
			\end{array}\right).
			\end{aligned}$$
			From the fact ${\rm rk}\left(\begin{array}{c}F\\R \end{array}\right)={\rm rk}(O(A,F))$, we have
			$${\rm rk}
			\left(\begin{array}{c}
			F\\
			R
			\end{array}\right)  \le \begin{aligned}{\rm rk}
			\left(\begin{array}{c}
			FA \\
			RA \\
			F\\
			R
			\end{array}\right)\le {\rm rk}(O(A,F))={\rm rk}\left(\begin{array}{c}
			F\\
			R
			\end{array}\right).
			\end{aligned}$$
			That is, condition \eqref{rank-condition} holds for $R$.
		\end{proof}
		\begin{remark} \label{rk3}
			Theorem \ref{thlower} reveals that the minimum (possible) order of augmented target output controllers is ${\rm rk}(O(A,F))$. This is consistent with Theorem \ref{th1}, noting that \eqref{modetest1} implies $r={\rm rk}(F)={\rm rk}(O(A,F))$.
			Let $v_1,...,v_r$ be the observability indices of $(A,F)$ corresponding to rows $F_1,...,F_r$ of $F$ (see \cite[Chap 6.2.1]{chen1984linear} for the definition of observability indices), where $F_i$ is the $i$th row of $F$, and let $O_i=\left(\begin{array}{c}F_iA\\
			\vdots\\
			F_iA^{v_i-1}
			\end{array}\right)$, $i=1,...,r$. As pointed out in \citep[Theo 6]{fernando2025existence},  a matrix $R$ with the minimum number of rows satisfying \eqref{rank-condition} can be constructed as
			$R=\left(\begin{array}{c}O_1\\
			\vdots\\
			O_r
			\end{array}\right)$. In this way, there is no need to compute the complete observability matrix $O(A,F)$ for constructing $R$.
		\end{remark}
		
		For system \eqref{sys1}, define a matrix $D_t$ as 
		\begin{equation*}
		D_t=\left(\begin{array}{cccccc}
		0 & 0 & 0 & \cdots & 0 & 0 \\
		F B & 0 & 0 & \cdots & 0 & 0 \\
		F A B & F B & 0 & \cdots & 0 & 0 \\
		\vdots & \vdots & \vdots & \ddots & \vdots & \vdots \\
		F A^{t-2} B & F A^{t-3} B & F A^{t-4} B & \cdots & F B & 0
		\end{array}\right),
		\end{equation*}
		Let ${\rm lag}(A,F)$ be the observability index of $(A,F)$, namely,
		$${\rm lag}(A,F)=\min\{\mu\in {\mathbb N}:{\rm rk}(O_{\mu}(A,F))={\rm rk}(O_{\mu+1}(A,F))\}.$$
		
		\begin{assumption}\label{asp2}
			A collection of experiment data $U=\{u_d(t)\}_{t=0}^T, X=\{x_d(t)\}_{t=0}^T$ and $Z=\{z_d(t)\}_{t=0}^T$ is available such that $\left(\begin{array}{c}U_{0, t, T-t+1} \\ X_{0, T-t+1}\end{array}\right)$ has full row rank for some $t\ge {\rm lag}(A,F)$.
		\end{assumption}
		
		\begin{remark}Since $\left(\begin{array}{c}U_{0, T-t+1} \\ X_{0, T-t+1}\end{array}\right)$ is a submatrix of $\left(\begin{array}{c}U_{p} \\ X_{p}\end{array}\right)\doteq \left(\begin{array}{c}U_{0, T} \\ X_{0, T}\end{array}\right)$,
			it is readily seen that Assumption \ref{asp2} implies Assumption \ref{as1}. When ${\rm lag}(A,F)$ is not known apriori, we can take $t$ to be its upper bound $n$. Note that the length $T$ of the historical data should satisfy $T\ge (m+1)t+n-1$. If system \eqref{sys1} is controllable and the input $U_{0:T}$ is persistently exciting of order $n+t$, then Assumption \ref{asp2} is
			guaranteed~\cite{van2020willems}.
		\end{remark}
		
		From \eqref{sys1}, we have
		\begin{equation}\label{sys-data}Z_{0, t, T-t+1}=O_t(A,F)X_{0, T-t+1}+D_tU_{0, t, T-t+1}.\end{equation}
		Let $\Gamma^{\perp}_{U}=I-U_{0,t,T-t+1}^-U_{0,t,T-t+1}$.
		Since $U_{0, t, T-t+1}\Gamma^{\perp}_{U}=0$, from \eqref{sys-data}, we derive
		\begin{equation}\label{sys-null}Z_{0, t, T-t+1}\Gamma^{\perp}_{U}=O_t(A,F)X_{0, T-t+1}\Gamma^{\perp}_{U}.\end{equation}
		By \citep[Lem 9.1]{verhaegen2007filtering}, it holds that $X_{0, T-t+1}\Gamma^{\perp}_{U}$ has full row rank under Assumption \ref{asp2}.
		Therefore, Lemma \ref{lem1} yields
		\begin{equation}\label{rank-lower} {\rm rk}(O_t(A,F))={\rm rk}(Z_{0, t, T-t+1}\Gamma^{\perp}_{U}).\end{equation}
		Moreover, from \eqref{sys-null}, we have that under Assumption \ref{asp2},
		\begin{equation}\label{ot} O_t(A,F)=Z_{0, t, T-t+1}\Gamma^{\perp}_{U}(X_{0, T-t+1}\Gamma^{\perp}_{U})^-.\end{equation}
		Based on \eqref{rank-lower} and \eqref{ot}, we provide a data-driven design for matrices $R$ with the minimum number of rows satisfying \eqref{datatest10aug}.
		
		\begin{theorem}[Data-Driven Design of Matrices $R$] \label{design-augmentation}
			Under Assumption \ref{asp2}, the minimum number of rows of matrices $R$ satisfying \eqref{datatest10aug}, given by $d_{\min}$, can be expressed as
			$$d_{\min}={\rm rk}(Z_{0, t, T-t+1}\Gamma^{\perp}_{U})-r.$$
			Moreover, let $R$ consist of $d_{\min}$ rows of $Z_{0, t, T-t+1}\Gamma^{\perp}_{U}(X_{0, T-t+1}\Gamma^{\perp}_{U})^-$ such that
			$${\rm rk}\left(\begin{array}{c} R\\ Z_pX_p^-\end{array}\right)={\rm rk}(Z_{0, t, T-t+1}\Gamma^{\perp}_{U}).$$
			Then, $R$ is a matrix with $d_{\min}$ rows satisfying \eqref{datatest10aug}.
		\end{theorem}

		\begin{proof} Note that $Z_pX_p^-$ gives $F$ from \eqref{linear-relation}. Based on the facts that Assumption \ref{asp2} implies Assumption \ref{as1} and $R$ satisfies \eqref{datatest10aug} if and only if it satisfies \eqref{rank-condition}, the required result follows from Theorem \ref{thlower}, \eqref{rank-lower}, and \eqref{ot}.
		\end{proof}
		
		Based on Theorem \ref{design-augmentation}, a data-driven procedure for designing a minimum-order augmented target output controller is given as follows: 
		\begin{itemize}
			\item 1. Check if condition \eqref{modetest5} holds (c.f. condition (3) of Theorem \ref{th4}). If yes, continue to step 2, otherwise stop.
			\item 2. Construct matrix $R$ according to Theorem \ref{design-augmentation}. 
			\item 3. Determine $T_1,T_2$ such that $\binom{Z_f}{RX_f}=T_1U_p+T_2\binom{Z_p}{RX_p}$. Check if $(T_2,T_1)$ is controllable.  If No, exit the algorithm. If YES, continue to step 4.
			\item 4. Determine $K$ such that $T_1K+T_2$ has a desired set of poles. Then the target output controller is $u(t)=K\binom{F}{R}x(t)$, with $F=Z_pX_p^-$. 
		\end{itemize}

	\section{Observer-based target output controllers}
	In the practical implementation of target output controllers, the whole state $x(t)$ may be unmeasurable. In this case, we can use an observer to estimate $x(t)$. We present this extension in this section, revealing a separation principle between the target output controller design and the observer design in the observer-based target output controllers.
	
	The observer-based target output controller is
	\begin{align}
	x(t+1) &= A x(t) + B u(t) \label{eq:1} \\
	u(t) &= K F \hat{x}(t) \label{eq:2} \\
	\hat{x}(t+1) &= A \hat{x}(t) + B u(t) + L \left(y(t) - C \hat{x}(t)\right) \label{eq:3}
	\end{align}
	where \eqref{eq:3} is a Luenberger observer that gives   an estimation $\hat x(t)$ of $x(t)$, where $L$ is the corresponding observer gain.
	
	\begin{theorem} \label{th6}
		Suppose that conditions \eqref{modetest1}-\eqref{modetest2} hold and $(A,C)$ is observable. Let $e(t)=x(t)-\hat x(t)$ and $N$ is such that $FA+FBKF=NF$. The closed-loop dynamics of the subsystem involving $Fx(t)$ is
		\begin{equation} \label{separation}
		\left(\begin{array}{c}
		Fx(t+1)\\
		e(t+1)
		\end{array}\right)=\left(\begin{array}{ll} N & -FBKF\\
		0 & A-LC \end{array}  \right)\left( \begin{array}{c}Fx(t) \\ e(t) \end{array} \right).
		\end{equation}
	\end{theorem}
	\begin{proof}
		Subtracting \eqref{eq:3} from \eqref{eq:1}, we get
		$$e(t+1)=(A-LC)e(t).$$
		Pre-multiplying $F$ and substituting $u(t)=KF\hat x(t)$ into \eqref{eq:1}, we get
		$$Fx(t+1)=NFx(t)-FBKFe(t),$$where $F(A+BKF)=NF$ and $\hat x(t)=x(t)-e(t)$ have been used.
		Equation \eqref{separation} then follows.
	\end{proof}

	Theorem \ref{th6} reveals a separation principle in the design of observer-based target output controllers. That is, the poles of the $(n+r)$-dimensional subsystem involving $Fx(t)$ consist of poles of the observer (eigenvalues of $A-LC$) and poles of the target output controller (eigenvalues of $N$), which can be designed separately and independently by choosing suitable $K$ and $L$.    Provided that the conditions in Theorem \ref{th6} hold, both pole sets can be designed arbitrarily.
	Based on the data-driven observers proposed in \cite{Turan2021} and the data-driven target output controllers in Section \ref{sec-3}, we can design a data-driven observer-based target output controller, where the observer and controller can be designed separately.
	
	More specific, under Assumption \ref{as1} on the historical data $(U,Y,X)$, find matrices $
	\big(\Sigma_{U_p}\ \Sigma_{Y_p}\ \Sigma_{Y_f}\ \Sigma_{X_p}\big)
	$ with $\Sigma_{X_p}$ Schur stable, such that

	\begin{equation}\label{eq:darouach}
	X_f = \left(\Sigma_{U_p} \ \Sigma_{Y_p} \ \Sigma_{Y_f} \ \Sigma_{X_p}\right)
	\left(\begin{array}{c} U_p \\[1mm] Y_p \\[1mm] Y_f \\[1mm] X_p \end{array}\right).
	\end{equation}
	Then, according to \cite{Turan2021,disaro2024equivalence}, a state observer can be constructed as
	\begin{equation}\label{eq:observer_update_main}
	\hat{x}(t+1) = \Sigma_{U_p}\,u(t) + \Sigma_{Y_p}\,y(t) + \Sigma_{Y_f}\,y(t+1) + \Sigma_{X_p}\,\hat{x}(t).
	\end{equation}
	Here, $\hat x(t)$ is an asymptotic estimation of $x(t)$, i.e., $x(t)-\hat x(t)\to 0$ as $t\to \infty$, and the error $e(t)=x(t)-\hat x(t)$ can converge to zero arbitrarily fast by designing the eigenvalues of $\Sigma_{X_p}$, provided that $(A,C)$ is observable. Let $K$ be designed according to Theorem \ref{th4}. Then, the observer-based control law is
	$$u(t)=KZ_pX_p^-\hat x(t),$$
	where $Z_pX_p^-$ gives $F$.

	
	\section{Numerical simulations}
	Consider system \eqref{sys1} with the following parameters, taken from ~\cite{fernando2025existence}
	$$
	\begin{gathered}
	A=\left(\begin{array}{rrrrr}
	1 & 0.5 & -1 & 0 & 1 \\
	0.3 & 0.5 & -0.6 & -0.3 & 0.3 \\
	-0.6 & 0 & 0.2 & 0.6 & -0.6 \\
	1.25 & 0.5 & -1 & -0.25 & 1.75 \\
	-0.75 & 0 & 0 & 0.75 & -0.25
	\end{array}\right), \\
	B=\left(\begin{array}{rr}
	1 & -1 \\
	1 & 1 \\
	0 & 0 \\
	1 & 0 \\
	0 & 1
	\end{array}\right), C=\left(\begin{array}{lllll}
	0 & 0 & 2 & 1 & 0 \\
	0 & 0 & 0 & 0 & 1
	\end{array}\right). \\
	\end{gathered}
	$$
	Since $(A,B)$ is not controllable (the dimension of its controllable subspace is $3$), arbitrary assignment of poles of the whole system by state feedback is impossible.
	In what follows, we present data-driven target output controller designs in two cases.
	
	{\emph{Case 1: Target output controller design by placing $r$ poles.}} Consider $$
	F=\left(\begin{array}{lllll}
	1 & 1 & -2 & 0 & 2
	\end{array}\right) .
	$$
	We generate historical data by
	applying a random input sequence of length $T = 19$ with entries drawn from the standard uniform distribution $(0,1)$  to the above system with random initial states.  It turns out that Assumption \ref{as1} is always satisfied for the obtained data.  We can verify that conditions \eqref{modetest4}-\eqref{modetest5} in Theorem~\ref{th2} hold, indicating that a data-driven $r$-order target output controller exists, $r=1$. We now follow the data-driven target output controller design algorithm in Theorem \ref{th5}, obtaining
	$$T_1=\left(\begin{array}{ll}
	2 & 2
	\end{array}\right), \ T_2=1.$$By choosing $K$ to be
	$$
	K=\binom{-0.1525}{-0.1525},
	$$
	the eigenvalue of $T_1K+T_2$ becomes $0.39$, which is Schur stable. In designing the observer \eqref{eq:observer_update_main}, $\Sigma_{X_p}$ is chosen to be
	{\small$$\Sigma_{X_p}\!=\!
		\left(\begin{array}{ccccc}
		0.4599 & 0.1551 & -0.1455 & 0.0707 & 0.2261 \\
		0.4004 & 0.3277 & 0.1173 & -0.4898 & 0.0748 \\
		-0.3988 & -0.0170 & -0.0482 & 0.2306 & -0.2755 \\
		0.8562 & 0.1759 & -0.0356 & -0.2166 & 0.6365 \\
		-0.2571 & -0.0295 & -0.0542 & 0.1532 & -0.0586
		\end{array}\right),
		$$}whose eigenvalues are $\{-0.5378, 0.4233 \pm 0.0990i, 0.0304, 0.1249\}$. The responses of the target output controllers based on the exact $x(t)$ and the observer-based estimation $\hat x(t)$ are shown in Fig.~\ref{figexample2}, where the corresponding initial state $x(0)$ and initial estimation $\hat x(0)$ are randomly selected.  While both trajectories are stabilized to zero, the observer-based trajectory has a bigger overshoot and a larger settling time. This is reasonable, since the observer-based target output controller uses an asymptotic estimation $\hat x(t)$ of $x(t)$.


	
	\begin{figure}[H]
		\centering
		\includegraphics[width=0.45\textwidth]{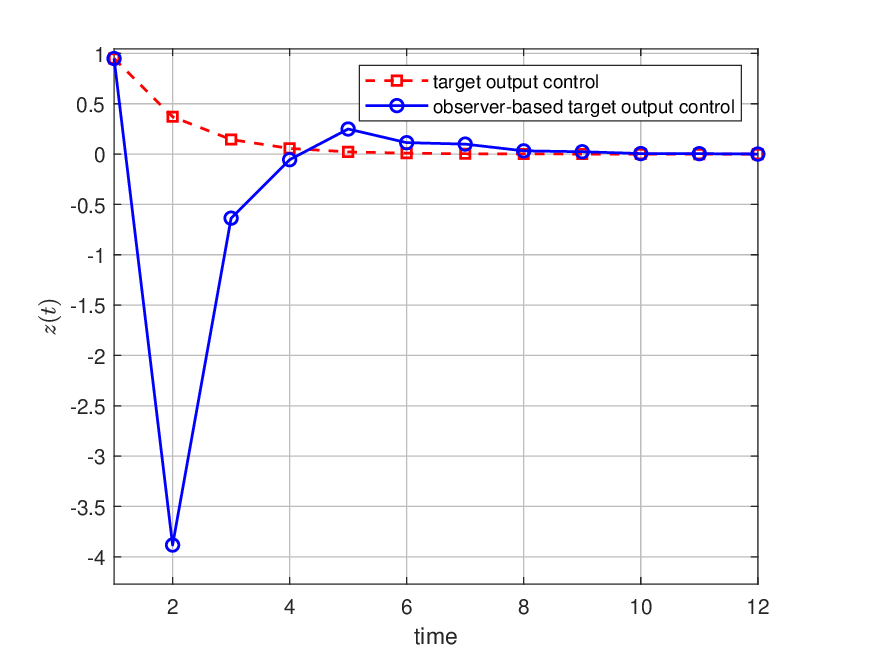}
		\caption{Responses of target outputs for Case 1.}
		\label{figexample2}
	\end{figure}	
	
	{\emph{Case 2: Target output controller design by target augmentation.}} Consider \begin{equation} \label{F-matrix}
	F=\left(\begin{array}{lllll}
	0.5 & 1 & -2 & 0.5 & 2.5
	\end{array}\right).
	\end{equation}Data generation follows the same procedure as in Case 1.  It turns out that conditions \eqref{modetest4}-\eqref{modetest5} in Theorem~\ref{th2} do not hold. Therefore, we need to augment the target outputs. {The procedure in Theorem \ref{design-augmentation} yields $d_{\min}=1$ and
		$$R=\left(\begin{array}{lllll}
		0.75 & 1 & -2 & 0.25 & 2.25
		\end{array}\right),$$which satisfies condition \eqref{datatest10aug} and also condition \eqref{datatest5aug}.} This yields an augmented target output controller of the minimum order $2$. The trajectories of the original target output $z(t)$ and the augmented output $Rx(t)$ with random initial conditions are presented in Fig. \ref{figexample3}, where both target output controllers based on the exact state and the observer-based state estimation are included. It turns out that
	the original target outputs and the augmented outputs are all regulated to zero.
	\begin{figure}[H]
		\centering
		\includegraphics[width=0.45\textwidth]{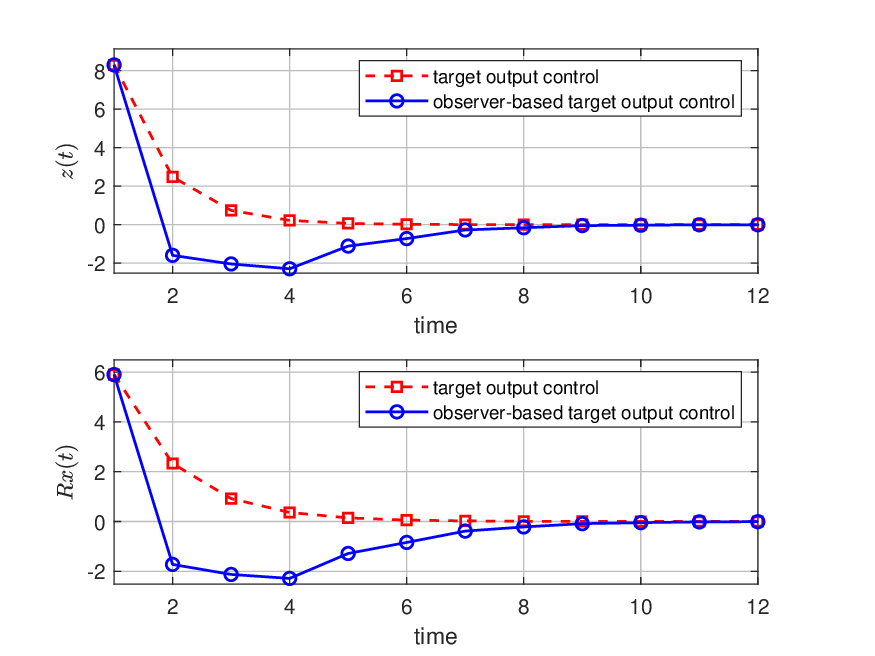}
		\caption{Responses of the target output controller in Case 2.}
		\label{figexample3}
	\end{figure}	

	\section{Conclusion} \label{section 5}
	
	This paper has addressed the existence and design of target output controllers in a data-driven framework, focusing on systems where full state controllability is not guaranteed. We have derived necessary and sufficient conditions for the existence of target output controllers based on input and partial state data, eliminating the need for explicit system parameter identification. Additionally, we have introduced data-driven methods for designing both standard and (minimum-order) augmented target output controllers. We further explored a separation principle between the design of target output controllers and state observers, leading to the development of observer-based target output controllers. Numerical simulations demonstrated the effectiveness of the proposed methods.
	Future work could incorporate functional observers \cite{darouach2000existence,fernando2010functional,zhang2023functional} in the design of observer-based target output controllers, thus eliminating the observability condition, and investigate the robustness of the proposed controllers with noisy data.
	
	
	\bibliographystyle{elsarticle-num}
	{\footnotesize
		\bibliography{yuanz3}
	}

\end{document}